\documentclass[prd,reprint,nofootinbib,amsmath]{revtex4-2}
\usepackage[dvips,final]{graphicx}
\usepackage{amssymb}
\usepackage{bm}
\usepackage{color}
\newcommand{\ba}{\begin{eqnarray}}                        
\newcommand{\ea}{\end{eqnarray}}                          
\newcommand{\be}{\begin{equation}}\newcommand{\ee}{\end{equation}}%
  \def\ie{\hbox{\it i.e.}{ }} 
   \def\eg{\hbox{\it e.g.}{ }} 
\begin{document}

\title{Endpoint behavior of distribution amplitudes of pion and longitudinally polarized rho meson under the influence of renormalon-chain contributions}

\author{S.~V.~Mikhailov}
\email{mikhs@theor.jinr.ru}
\author{N.~Volchanskiy}
\affiliation{Bogoliubov Laboratory of Theoretical Physics, Joint Institute for Nuclear Research, 141980 Dubna, Russia}
\date{\today}

\begin{abstract}
We calculate the two-point massless QCD correlator of nonlocal (composite) vector quark currents with chains of fermion one-loop radiative corrections inserted into gluon lines.
The correlator depends on the Bjorken fraction $x$ related to the composite current and, under large-$\beta_0$ approximation, gives the main contributions in each order of perturbation theory.
In the mentioned approximation, these contributions dominate the endpoint behavior of the leading-twist distribution amplitudes of light mesons in the framework of QCD sum rules.
Based on this, we analyze the endpoint behavior of these distribution amplitudes for $\pi$ and longitudinally polarized $\rho^\|$  mesons and find inequalities for their moments.
\end{abstract}

\maketitle


\section{Introduction}

Distribution amplitudes (DA) of mesons are the key hadronic characteristics in hard exclusive reactions with participation of hadrons -- due to ``factorization theorems'', they determine the behavior of the form factors and amplitudes of the corresponding exclusive processes.
The DA reflects the consequences of the long-distance QCD dynamics for partons within the meson that carries the $xp$  fraction of the
meson momentum $p$.
Here we investigate the role of higher radiative corrections to the correlator of nonlocal currents in its relation to QCD sum rules (SR)
for DAs of light mesons.
Finally, we will focus on the behavior of DAs near the endpoints of the interval $x \in (0,1)$.
We have two main, different in nature, radiative contributions to QCD SR for the $\pi/\rho$-meson (light meson)
DAs \cite{Bakulev:1998pf,Mikhailov:2021kty} that determine the behavior near the endpoints:
(i) $\alpha_s$-corrections to the purely perturbative part of the corresponding correlator \cite{Mikhailov:2020izb,Mikhailov:2023lqe},
(ii) and $\alpha_s$-corrections to the four-quark condensate interaction for this correlator.
Both kinds of corrections are considered here in the renormalon $n$-bubble approximation
to massless perturbative QCD\footnote{In QCD, SU$(N_c)$ with $N_c=3$, the Casimir invariants are $C_A=N_c$ and $C_F=T_R(N_c^2-1)/N_c$, $T_R = 1/2$. The one-loop $\beta$-function coefficient is $\beta_0 = \frac{11}{3} C_A - \frac43 T_R n_f = 9$ at $n_f=3$ massless quark flavors. $a_s = \alpha_s /(4\pi)$ is the coupling constant.}.

The paper is organized as follows. In Sec.~\ref{sec:2}, we start with the results of calculating the two-point correlators $\Pi_{n}(x,\underline{0};L)$ of nonlocal vector quark currents within the $n$-bubble approximation
(or, equivalently, at large $\beta_0 a_s$) in $\overline{\text{MS}}$ scheme,
\begin{align}\label{eq:cor-def}
-i\frac{a_s}{\pi^2} N_c & C_F A^n \Pi_{n}(x,y;L)
\notag\\
={}& \int \mathrm{d}^D\eta \, e^{ip\eta} \langle 0|\hat{\mathrm{T}}\left[J^\dagger(\eta;x)J(0;y) \right]|0 \rangle
\notag\\
 ={}& \vcenter{\hbox{\includegraphics[width=.13\textwidth]{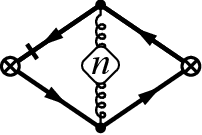}}}
 + \vcenter{\hbox{\includegraphics[width=.13\textwidth]{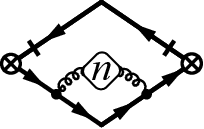}}}
\notag\\
 &{}
 + \vcenter{\hbox{\includegraphics[width=.13\textwidth]{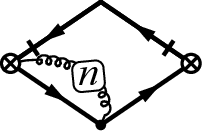}}}
 + \vcenter{\hbox{\includegraphics[width=.13\textwidth]{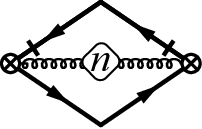}}}
 + \dots
\end{align}
Here, $\eta$ is a space-time point; $L=\ln(P^2/\mu^2)$ with $P^2=-p^2$,~$p$ being the external momentum and $\mu$ the renormalization scale,
and the constant $ A = \frac43 a_s T_F n_f$ can be replaced by $-a_s \beta_0$ as is prescribed by the naive nonabelianization (NNA) trick \cite{Broadhurst:1994se}, where $a_s = \alpha_s /(4\pi)$.
In Eq.~\eqref{eq:cor-def}, the nonlocal vector quark current $J(\eta;x)$, denoted diagrammatically with a vertex $ \pmb{\otimes}$,
is defined as the inverse Mellin transform $\hat{\mathtt{M}}^{-1}$ of a quark bilinear involving the $N$th derivative of the quark field operator:
\begin{align}\label{eq:currents-def1}
\!\!\!	J(\eta;x) = \hat{\mathtt{M}}^{-1}J(\eta;\underline{N}),
	J(\eta;\underline{N}) = \bar{d}(\eta) \hat{\tilde{n}} \left(i\tilde{n}\nabla\right)^N u(\eta),
\end{align}
where $x$ is the Bjorken fraction; $\nabla_\mu=\partial_\mu -i g t_a A^a_\mu$ is the QCD covariant derivative; $\tilde{n}^\mu$ is the light-like vector, $\tilde{n}^2=0$. In Eqs.~\eqref{eq:cor-def} and \eqref{eq:currents-def1}, as everywhere in what follows, the arguments of the Mellin transform are underlined, i.e., $f(\underline{a}) = \hat{\mathtt{M}} f(x) = \int_0^1 \mathrm{d}x \, f(x) x^a $.
The nonlocal current \eqref{eq:currents-def1} emerges naturally in the description of QCD factorization in hard exclusive processes---its projection on a meson state gives the corresponding DA of the leading twist.
The investigation of the correlator $\Pi_{n}(x,y;L)$ is a general problem consisting of a few parts that will be a subject of another publication.
Below, we consider a special case of the correlator $\Pi_n(x,\underline{0};L)$ and its derivatives with respect to $L$
which have an immediate application in the area of QCD SRs, as discussed in Sec.~\ref{sec:SRs}.


\section{\label{sec:2}Correlator $\Pi_n(x,\underline{0};L)$ and  QCD SR}

Within the approach of QCD SR, the Borel transform $\hat{\mathbf{B}}$ (see discussion in Appendix \ref{sec:Borel}) of the correlator~\eqref{eq:cor-def} determines the perturbative contributions to meson DAs,
\begin{gather}\label{eq:DA-def}
\!\!\!\!\text{DA}(x;l)\! \sim \!\frac{ N_c}{\pi^2}\Big[ \frac12 x \bar{x}+ a_s C_F\hat{\mathbf{B}}\sum_{n \geqslant 0} A^n \Pi_{n}(x,\underline{0};L)\Big],
\\
\Pi_n(x,\underline{0};L)=\int_0^1 \Pi_n(x,y;L) dy,
\end{gather}
where $\bar{x} =1-x$ and $l = \ln(M^2/\mu^2)$ is the logarithm of the Borel parameter $M^2$ appearing in the Borel transform $\hat{\mathbf{B}}[L^m]$ of
the powers of $L=\ln(P^2/\mu^2)$, see Appendix~\ref{sec:Borel}.
In the approximation of large $\beta_0$ (or $n_f$), the pQCD part of SR is completely determined by the diagrams \eqref{eq:cor-def}
of the two-loop topology with gluon lines dressed by chains of one-loop fermion bubbles---renormalon chains
\begin{gather*}
\vcenter{\hbox{\includegraphics[width=.10\textwidth]{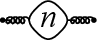}}}
 = \underbrace{\vcenter{\hbox{\includegraphics[width=.34\textwidth]{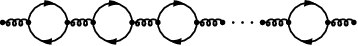}}}}_{n}.
\end{gather*}
The general expression for these  diagrams of the two-loop topology  with nonlocal vertices and an arbitrary exponent of the gluon line propagator was derived in \cite{Mikhailov:2018udp}.
This ``kite'' diagram can be represented in terms of the hypergeometric functions $_3F_2(x)$ and $_3F_2(\bar{x})$,
which we will meet below in the generating functions for $\Pi_{n}(x,\underline{0};L)$ \cite{Mikhailov:2023lqe}.


\subsection{\label{sec:x0-correlator}The generating function for the correlator $\Pi_n(x,\underline{0};L)$}

Here we  briefly  discuss the properties of $\Pi_n(x,\underline{0};L)$,
which is the two-point $n$-bubble correlator of one local and one nonlocal (dependent on the Bjorken fraction $x$) quark current,
as defined in Eq.~\eqref{eq:cor-def}.
The sequence $\Pi_n(x,\underline{0};L)$ can be split into two parts originating from expansions of different generating functions,
exponential $\Pi_{n}'$ and ordinary $\Pi_{n}''$, see \cite{Mikhailov:2023lqe}:
\begin{align}\label{eq:pix0}
\Pi_{n}(x,\underline{0};L) = \Pi_{n}'(x,\underline{0};L) + \Pi_{n}''(x,\underline{0};L).
\end{align}
Further, we will consider two quantities derived from $\Pi_{n}$ --- its Borel image that is defined in \eqref{eq:borel}, $\hat{\mathbf{B}}[\Pi_{n}]$,
and the first derivative $\displaystyle \dot{\Pi}_{n}\equiv \frac{d}{dL}\Pi_{n}$,
the later is useful for comparison with the known results for the Adler $D$ function:
\begin{align}\label{eq:pix0'}
&\sum_{n \geqslant 0} \frac{A^n}{n!}\dot{\Pi}_{n}'(x,\underline{0};L) = \frac{e^{A\, L_c} }{A^2(1+A)(2+A)}
\notag \\
& {} \times {}\mathop{\hat{\mathbf{S}}} \Biggl\{ x^A \Biggl[ -\bar{x} (A + 4 x ) + 2 x \bar{x} \frac{(\pi A)^2 \cot(\pi A)}{x^A\sin(\pi A)}
\notag \\ &  \hphantom{ {} \times {}\mathop{\hat{\mathbf{S}}} \Biggl\{ x^A \Biggl[  }
{} + x (2\bar{x}+A) A \text{B}_{\bar{x}}(A,1-A)
\notag \\ &  \hphantom{ {} \times {}\mathop{\hat{\mathbf{S}}} \Biggl\{ x^A \Biggl[  }
{} + \frac{2x^2\bar{x}A^2}{(1+A)^2} {}_3F_2  \left(  \left.\begin{matrix}
																						1,\, 1,\, 1+A \, \\
																						2+A,\, 2+A \,
																	\end{matrix} \right\rvert x \right)
 \Biggr] \Biggr\},
\\
& \sum_{n \geqslant 0} A^n \dot{\Pi}_{n}''(x,\underline{0};L)
\notag\\ & \hphantom{ \sum_{n \geqslant 0} A^n  }
{}= - \frac1{2 A} \int_0^A \frac{\mathrm{d}a }{a}\int_0^1 \left[ \frac{ V(x,y;a)}{ h_1(a) }\right]_{+(x)} y\bar{y}\, \mathrm{d}y,\label{eq:pix0''}
\end{align}
where
\begin{subequations}
\begin{gather}
h_1(a) = \frac{(1-a) \Gamma(1+a) \Gamma^3(1-a)}{(1-2a/3) (1-2a) \Gamma(1-2a)},\quad L_c = L-5/3, \label{eq:2.5a}
\\
V(x,y;a) = 2 \mathop{\hat{\mathbf{S}}} \left[ \theta(y>x) \left( \frac{x}{y} \right)^{1-a} \left( 1-a + \frac{1}{y-x} \right) \right].
\end{gather}
\end{subequations}
Here, $h_1(\varepsilon)$ comes from the $\varepsilon$-dependence of the fermion one-loop correction on the gluon propagator ($D=4-2\varepsilon$ is the space-time dimension), its expansion in $\varepsilon$ in the first order leads to the shift $c=-5/3$ in $L_c$;
$V(x,y;a)$ is a generalization \cite{Mikhailov:1998xi,Mikhailov:1999ht}
of the one-loop ERBL evolution kernel that allows one to take into account renormalon-chain corrections to $V_0(x,y)=V(x,y;0)$; $f(x,y)_{+(x)} = f(x,y) - \delta(x-y) f(\underline{0},y)$ is the plus distribution; $\mathop{\hat{\mathbf{S}}} \left[ f(x,y) \right] = f(x,y) + f(\bar x, \bar y)$.
The part $\Pi''$ in \eqref{eq:pix0''} that is obtained from the ordinary generating function is completely determined by the counterterms to the nonlocal vertex.
From \eqref{eq:pix0}--\eqref{eq:pix0''}, we can derive explicit coefficients of the $L$-expansion of the correlator
\begin{gather}
\Pi_n(x,\underline{0};L) = (-1)^n n! \sum_{k=0}^{n+1} \frac{(-L)^k}{k!} \Pi_n^k(x,\underline{0}).
\end{gather}
The highest degree term $\Pi_n^{n+2}(x,\underline{0})$ is proportional to $\int_0^1 V_0(x,y)_{+} y\bar{y}\,dy =0$  due to the vector current conservation.
The first nonvanishing coefficient at $k=n+1$ reads
\begin{align}
\Pi&_{n}^{n+1}(x,\underline{0}) = \frac12 \mathop{\hat{\mathbf{S}}}  \Biggl\{ x \ln x
\notag\\
&{} - \delta_{0,n} \left[x \ln x - \frac12 x \bar{x} \left( \frac{\pi^2}{3} - 5 - \ln^2\left( \frac{x}{\bar{x}}\right) \right) \right] \Biggr\},
\end{align}
which is in agreement with the previous calculations for $n=0,1$ in \cite{Mikhailov:2020tta}.
The consequent terms are too cumbersome  to be written out here.
Nevertheless, the highest transcendence types of functions appearing in further orders
can be expressed in terms of (harmonic) polylogarithms \cite{Mikhailov:2023lqe}:
\begin{align*}
\Pi_{n > 0}^n(x,\underline{0}) &{} \propto \mathop{\hat{\mathbf{S}}} \mathop{\mathrm{Li}_{3}}(x) + \text{simpler polylogarithms},
\\
\Pi_{n > 1}^{n-1}(x,\underline{0}) &{}\propto \mathop{\hat{\mathbf{S}}} \mathop{\mathrm{Li}_{4}}(x) + \dots,
\\
\Pi_n^{k>0}(x,\underline{0}) &{}\propto \mathop{\hat{\mathbf{S}}} \mathrm{H}_\text{\bm{${\mu}$}} (x) + \dots,
\end{align*}
where $\mathrm{H}_\text{\bm{${\mu}$}} (x)$ are harmonic polylogarithms \cite{Remiddi:1999ew} with multi-index $\text{\bm{${\mu}$}}=\mu_1,\dots \mu_r: \;\mu_i>0,\; \sum \mu_i = n-k+3$.

\begin{figure}[t]
\begin{center}
\includegraphics[width=0.48\textwidth]{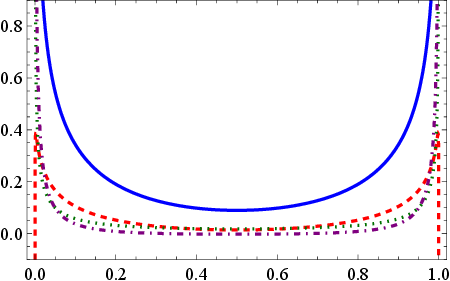}
\caption{The ratios of $\hat{\mathbf{B}}[\Pi_{n}]$ to the one-loop correlator $\text{LO}= x\bar{x}/2$: $a_s C_F \hat{\mathbf{B}}[\Pi_{0}]/$LO (solid blue line), $a_s^2 C_F  \beta_0\hat{\mathbf{B}}[\Pi_{1}]/$LO (dashed red line), $a_s C_F (a_s\beta_0)^2\hat{\mathbf{B}}[\Pi_{2}]/$LO (dotted green line), and $a_s C_F (a_s\beta_0)^3\hat{\mathbf{B}}[\Pi_{3}]/$LO (dash-dotted purple line).
All curves are for the case of $l=0$, $\alpha_s (\mu^2 = 1 \text{ GeV}^2) \approx 0.49$. \label{fig:x0}}
\end{center}
\end{figure}%

Figure~\ref{fig:x0} shows several lowest-order contributions to meson DAs in Eq.~(\ref{eq:DA-def}) given by the Borel transform \eqref{eq:borel} of Eqs.~\eqref{eq:pix0}--\eqref{eq:pix0''}. 
These curves exhibit different behaviors for the intermediate values of $x$, where they decrease sequentially from LO to N$^4$LO, and at the endpoints, where their ratios become singular. The vicinity of endpoints is quantitatively important for the form factors of the mesons considered.
Therefore, it makes sense to look at two integral characteristics of the correlators $\hat{\mathbf{B}}[\Pi_{n}(x,\underline{0};L)]$, their zeroth $\hat{\mathbf{B}}[\Pi_{n}(\underline{0},\underline{0};L)]$
and inverse $\hat{\mathbf{B}}[\Pi_{n}(\underline{-1},\underline{0};L)]$ moments.
They are mostly influenced by intermediate and near-endpoint values of the $x$-dependent correlator, respectively.


\subsection{\label{sec:2.2}The zeroth moment $\Pi_{n}(\underline{0},\underline{0};L)$.}

\begin{figure}[ht]
\begin{center}
\includegraphics[width=0.48\textwidth]{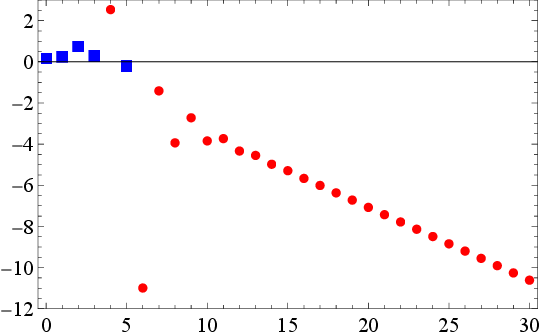}
\includegraphics[width=0.48\textwidth]{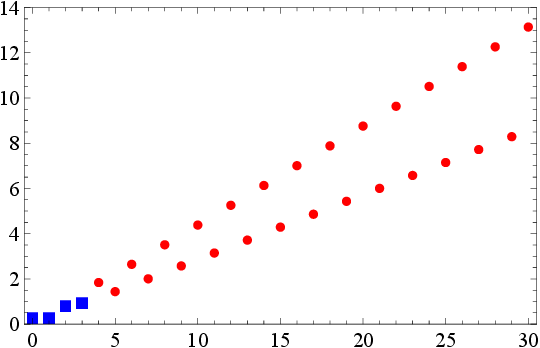}
\caption{\label{fig:moms}The ratio $R_n(\underline{N}) = -a_s \beta_0 \hat{\mathbf{B}}[ \Pi_n(\underline{N},\underline{0};L)] / \hat{\mathbf{B}}[ \Pi_{n-1}(\underline{N},\underline{0};L)]$. Top: $N=0$. Bottom: $N=-1$.
$R_0$ is defined as the ratio of 2-loop and 1-loop correlators, $R_0(\underline{0}) = 3 a_s C_F$ and $R_0(\underline{-1}) = 5 a_s C_F$ \cite{Mikhailov:1988nz, Mikhailov:2020tta}. Blue squares are for $R_n \leqslant 1$. All free parameters are the same as in Fig.~\ref{fig:x0} }
\end{center}
\end{figure}

The derivative of the zeroth moment
$\dot{\Pi}_{n}(\underline{0},\underline{0};L)$
is proportional to the Adler function $D(a_s)$ of QCD.
The corresponding exponential generating function $(A \to u )$ reads
\begin{align}
\mathbf{\tilde B}\dot{\Pi}(u) & {} \equiv\sum_{n \geqslant 0} \frac{u^n}{n!} \dot{\Pi}_{n}(\underline{0},\underline{0};L)
\notag\\
&{} = \frac{2e^{u L_c}}{3(1+u)(2+u)} \Big[ \Phi(-1,2,1-u)
\notag\\ & \hphantom{{} = \frac{2e^{u L_c}}{3(1+u)(2+u)} \Big[ }
{}- \Phi(-1,2,3+u) \Big]\label{eq:Pi(0,0)},
\end{align}
where the function $\Phi$ is Lerch's transcendent.
Using the identity
\begin{eqnarray}
 \Phi(-1,2,z)= \frac14 \left[\psi_1\left(z/2\right) - \psi_1\left((z+1)/2\right)\right],
\end{eqnarray}
where $\psi_1$ is the trigamma function, one can arrive at other forms for $\mathbf{\tilde B}\dot{\Pi}(u)$ \cite{Mikhailov:2023lqe, Laenen:2023hzu}.
Also, it coincides with the Adler function $D(a_s,L)$ from \cite{Ball:1995ni} for $n=2, \, 3$ and with the all-order result for $D(a_s,L)$ from \cite{1993ZPhyC..58..339B,Broadhurst:1993ru}.
The behavior of the Borel transform $\hat{\mathbf{B}}[\Pi_{n}(\underline{0},\underline{0};L)]$ is depicted in Fig.~\ref{fig:moms}.
This asymptotic series should be truncated at $n=3$ where it becomes divergent and bursts into factorial growth at $n \simeq 10$.
\subsection{\label{sec:2.3}The inverse moment $\Pi_{n}(\underline{-1},\underline{0};L)$}

The two generating functions for the inverse moment can be written \cite{Mikhailov:2023lqe} as
\begin{gather}
\Pi_{n}(\underline{-1},\underline{0};L) = \Pi'_n(\underline{-1},\underline{0};L)+\Pi''_n(\underline{-1},\underline{0};L),
\end{gather}
\begin{align}
\mathbf{\tilde B}\hat{\mathbf{B}}[\Pi](u)
&{} \equiv \sum_{n \geqslant 0} \frac{u^n}{n!} \hat{\mathbf{B}}[\Pi'_{n}(\underline{-1},\underline{0};L)]
\notag \\
&{} = \frac{-e^{u L_c}}{2 \Gamma(1-u)(1+u)(2+u)}
\notag \\
& \hphantom{ {}= }
{} \times \Bigl[ \psi_1\left(\frac{2-u}{2}\right) - \psi_1\left(\frac{1-u}{2}\right) \Bigr],
\end{align}
\begin{gather}
\sum_{n \geqslant 0} A^n \hat{\mathbf{B}}[\Pi''_{n}(\underline{-1},\underline{0};L)] = \frac{1}{A} \int_0^A \mathrm{d}a F(\underline{-1},a),
\end{gather}
where
\begin{align*}
F(\underline{-1},{}& a) = \frac{\Gamma(4-2a)}{6\Gamma(2-a)^2 \Gamma(3+a)}
\\ &{} \times \left\{ \frac{5+6a-5a^2}{\Gamma(3-a)} + \frac{(1+2a) [\psi(1-a)- \psi(1)]}{a \Gamma(1-a)}\right\}.
\end{align*}
Figure~\ref{fig:moms} illustrates the behavior of the sequence $\hat{\mathbf{B}}[\Pi_{n}(\underline{-1},\underline{0};L)]$ that can be obtained with the help of \eqref{eq:borel}. The series becomes factorially divergent at $n=4$.


\section{\label{sec:SRs}QCD SR for the $\pi$/$\rho^{\|}$ DA of the leading twist}
The QCD SRs for the pion and longitudinally polarized $\rho$-meson DAs of the leading twist 2, $\varphi_\pi$ and $\varphi^{\|}_\rho$,
respectively, read \cite{Bakulev:1998pf,Mikhailov:2021kty}
\begin{subequations}\label{eq:sr}
\begin{align}
&\left(f_{\pi}\right)^2\varphi_\pi(x)
+ \left(f_{A_1}\right)^2\varphi_{A_1}(x) e^{-m^2_{A_1}/M^2}
\notag \\
&\!=\! \Phi_\text{PT}(x;M;s_0^A) + \Phi_S(x;M) + \Delta_C \stackrel{\text{def}}{=}\! \Phi_\pi(x,M),
\label{eq:sr-pi} \\
& \left(f^{\|}_{\rho}\right)^2 \varphi^{\|}_{\rho}(x)e^{-m^2_\rho/M^2}
 + \left(f^{\|}_{\rho'}\right)^2\varphi^{\|}_{\rho'}(x) e^{-m^2_{\rho'}/M^2}
\notag \\
&\!=\! \Phi_\text{PT}(x;M;s_0^V) - \Phi_S(x;M) + \Delta_C \stackrel{\text{def}}{=}\! \Phi_{\rho}(x,M), \label{eq:sr-rho}
\end{align}
where  $\Phi_S(x;M)$ is the scalar-condensate contribution and
\begin{eqnarray}
\Phi_\text{PT}(x;M;s_0^A) = \int_{0}^{s_0^A} \rho_\text{pt}(x;s)e^{-s/M^2}ds,\, \label{eq:PT-rho}
\end{eqnarray}
\begin{align}\label{eq:Delta}
\Delta_C = \Delta\Phi_G(x;M) + \Delta\Phi_V(x;M) + \sum_{i=1}^{3} \Delta\Phi_{T_i}(x;M).
\end{align}
\end{subequations}

In Eqs.~\eqref{eq:sr}, $\varphi_{A_1}$ and $\varphi^{\|}_{\rho^{'}}$ are the DAs for the next resonances,
$s_0^A$ and $s_0^V$ are the duality intervals
 in the axial (for pion) and vector (for $\rho$ meson) channels, respectively.
Remarkably, the right-hand side (rhs)\ $\Phi_{\pi}$ and $\Phi_{\rho}$ of QCD SRs (\ref{eq:sr}) for these two channels differ only in sign of the scalar-condensate contribution $\Phi_S$.
The reason for that was discussed in \cite{Mikhailov:2021kty}.

\begin{figure}[h]
\begin{gather*} 
\!\mathbf{\hat{B}^2}_{(s \to P^2)}
\begin{gathered}\includegraphics[width=.09\textwidth,bb= 0in 0in .642in 0.49in]{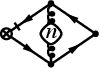}\end{gathered}
+\ldots \sim \rho_\text{pt}(s);\,
 \mathbf{\hat{B}}_{(M^2 \to P^2)}
\begin{gathered}\includegraphics[width=.09\textwidth,bb= 0in 0in .642in 0.49in]{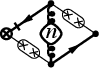}\end{gathered}
+\! \text{m.c.}
\end{gather*}
\caption{\label{fig:renorm+cond}Left diagram: the renormalon-chain contribution to the perturbative part of QCD SR, $ \Phi_\text{PT}$, via the density $\rho_\text{pt}$. Right diagram: the contribution of $a_s$-corrections to $\Phi_S(x;M)$ with a pair of nonlocal scalar condensates depicted by two ovals; the hard propagators of the coefficient function with a renormalon-chain are emphasized with thicker lines; m.c. here means
the mirror conjugate diagram.}
\end{figure}

The purely perturbative contributions $\Phi_\text{PT}(x;M;s_0^A)$ and $\Phi_\text{PT}(x;M;s_0^V)$ in the rhs\ of Eqs.~\eqref{eq:sr-pi} and \eqref{eq:sr-rho} can be obtained from higher order corrections to $\Pi(x;L)$ by integrating the spectral density $\rho_\text{pt}(x;s)= \mathbf{\hat{B}^2}_{(s \to P^2)}\Pi(x;L)$ in Eq.~(\ref{eq:PT-rho}), see Appendix~\ref{sec:rho}.
These perturbative terms dominate in the rhs\ of Eqs.~\eqref{eq:sr-pi} and \eqref{eq:sr-rho} in accordance with the standard practice of processing QCD SR \cite{Shifman:1978bx}.
The first two terms of $\rho_\text{pt}$ are $s$-independent and have been known \cite{PhysRevD.54.2182,Mikhailov:1988nz} for a long time,
\begin{align}
\rho_\text{pt}&(x;s)= \frac{N_c}{2\pi}
\notag\\
&{}\times\biggl( x\bar{x} +a_s C_\text{F} x\bar{x}\left[5-\frac{\pi^2}{3}+ \ln^2\left(\frac{\bar{x}}{x}\right) \right]+\ldots  \biggr).
\end{align}
In the vicinity of the endpoints $x=0$ and 1, the scalar condensate $\Phi_S(x;M)$ dominates the nonlocal condensate (NLC) contributions that include condensates $\Delta\Phi_{G,V,T_i}(x;M)$ \cite{Bakulev:1998pf,Bakulev:2001pa,Bakulev:2004mc} collected in the term $\Delta_C(x;M)$ in Eq.~(\ref{eq:Delta}). To estimate the behavior near the endpoints, we take into account only these \textit{two dominant terms}, $\Phi_\text{PT}$ and $\Phi_S$, in the rhs\ of Eqs.~(\ref{eq:sr-pi}) and \eqref{eq:sr-rho}, which is represented diagrammatically in Fig.~\ref{fig:renorm+cond}.

Note, that we apply  here the usual factorization approximation for the four-quark condensate.
Our estimates will be made under the renormalon-chain approximation for pQCD corrections, or,
in other words, in the approximation of large $\beta_0 a_s$ in both $\Phi_\text{PT}$ and $\Phi_S$.
Let us call ``reduced NLC SRs''  those that contain in their rhs\ of (\ref{eq:sr}) only the dominant
terms $\Phi_\text{PT}$, $\Phi_S$, while all other contributions are neglected.


\subsection{Effects of renormalon-chain corrections to pion DA}

\begin{figure}[h]
\includegraphics[width=.47\textwidth]{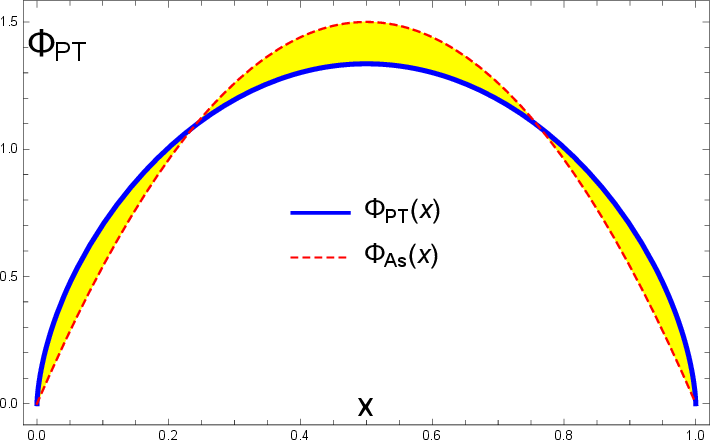}
\includegraphics[width=.47\textwidth]{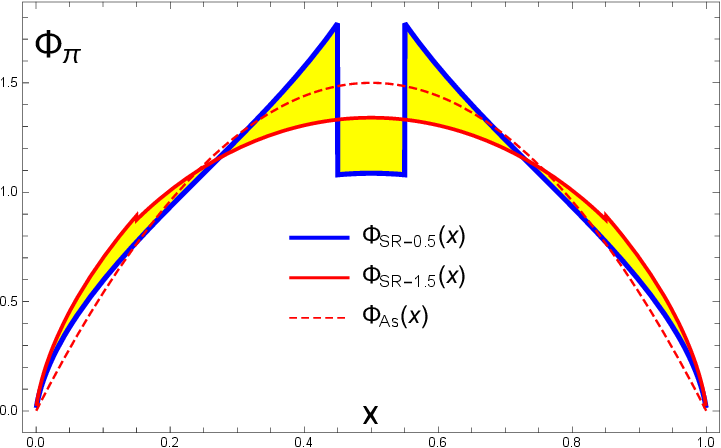}
\caption{\label{fig:qcdsr}Top: the perturbative part $\Phi_\text{PT}$ of the rhs\ of NLC SR up to $a_s(a_s\beta_0)^3$ (solid blue line) in comparison with $\Phi_\text{As}(x)$ (dashed red line). Bottom: the rhs\ of NLC SR, $\Phi_\pi$, is the sum of the condensate and perturbative contributions up to $a_s(a_s\beta_0)^3$ for the Borel parameter $M^2$ in the interval $[M^2_-=0.5$ (blue line), $M^2_+= 1.5$ (red line)] GeV$^2$ at its lower and upper bounds.}
\end{figure}

With growing  powers of $\beta_0 a_s$, the bubble-chain corrections lead to ``swelling'' of the perturbative part $\Phi_\text{PT}$ of DA $\varphi_\pi$ at the endpoints, which is shown in the top panel of Fig.~\ref{fig:qcdsr} in comparison with the leading-order contribution --- the asymptotic $\Phi_\text{As} = 6 x \bar{x}$. We restrict ourselves to considering orders up to $a_s(a_s\beta_0)^3$ (the next order $a_s (a_s\beta_0)^4$ does not change the result significantly) for which the series convergence stays good enough and the series does not yet succumb to factorial growth, see the discussion of Fig.\ref{fig:moms} in Secs~\ref{sec:2.2}--\ref{sec:2.3}.
At the same time, the corrections to the NLC part $\Phi_S(x;M)$ have the opposite effect, see top panel of Fig.~\ref{fig:qcdsr-rho} --- they alleviate the swelling of the perturbative part $\Phi_\text{PT}$.
The final result of this mutual compensation in the sum $\Phi_\text{PT}(x;M)+\Phi_S(x;M)$ is illustrated in Fig.~\ref{fig:qcdsr} (bottom panel).
We should mention here that Eqs.~(\ref{eq:sr}) should be considered as equalities \textit{in a weak sense}, i.e., for smooth convolutions of both sides of equations within the stability domain in $M^2$.
Usually, such convolutions are chosen as moments $ (\xi=2x-1)^N$ or $x^{-1}$, but, in general,
it can be any appropriate function of $x$.
In addition, the rhs\ of the QCD SR for DA should not be a smooth function of $x$, the smoothness of its behavior depends on a certain model for NLC, see, e.g.,
discussion in \cite{Mikhailov:2021kty}.
We use here the simplest Gaussian model for the NLC \cite{Mikhailov:1986be,Mikhailov:1988nz,Bakulev:1998pf}
that introduces a single parameter for nonperturbative QCD vacuum, an average virtuality of vacuum quarks
$\langle k_{q}^{2} \rangle = \lambda_{q}^{2} \equiv \langle\bar{q}D^{2}q\rangle /\langle\bar{q}q\rangle\Big|_{\mu_0^2\simeq 1\text{GeV}^2}$
at $\lambda^2_q \approx 0.45$ GeV$^2$ \cite{Mikhailov:2021znq}.
This model ignores any (still speculative) details of vacuum quark-gluon distributions at the cost of finite discontinuous contributions to the rhs\ of NLC SR, see the behavior of solid blue/red curves for $\Phi_S(x;M)$ in Fig.~\ref{fig:qcdsr-rho} (top). The contribution of $\Phi_S(x;M)$ is comparable to $\Phi_\text{PT}(x;M)$ near the
lower bound $M^2_-$ (blue curve) of the stability interval and significantly decreases at the 
upper bound $M^2_+$ (red curve). Let us briefly clarify the calculation of $\Phi_S(x)$ presented as a right diagram in Fig.~\ref{fig:renorm+cond}.
The $\Phi_S(x)$ is a convolution of a pair of scalar NLCs and a coefficient function (for details see \cite{Mikhailov:1988nz,Mikhailov:2010ud}), the latter includes now a renormalized bubble-chain.
Due to the Gaussian decay of the scalar NLC, the corresponding Feynman integrals for this convolution are well convergent and do not need to be renormalized.
An important calculation of $\Phi_ S(x)$ of the order $a_s(a_s\beta_0)^0$ was performed in \cite{Mikhailov:1988nz}
(see also Appendix~A of \cite{Mikhailov:2010ud}) --- our calculations  are similar to those.
The key integrals for the bubble-chain inclusion in the gluon line of the coefficient function are presented in Appendix~\ref{sec:app.C}.

Our goal here is to estimate how the QCD corrections affect the behavior of DA at the vicinity of endpoints rather than the whole profile of DA. 
Moreover, extending the analyses to moderate values of the Bjorken variable would require taking into account the other condensates $\Delta\Phi_{G,V,T_i}(x;M)$, which are numerically significant somewhere in the middle of the interval of $x$. So one can expect that the profile of the ``true'' pion DA lies somewhere within the yellow region between the blue (at $M^2=M^2_-$) and red (at $M^2=M^2_+$) bounds in Fig.~\ref{fig:qcdsr}(bottom) (with some uncertainty in the middle of the $x$-interval). The incline of DA near the endpoints varies\footnote{For this estimate we have used the technique of average incline elaborated in \cite{Mikhailov:2010ud}} from 6 to 7.
The inverse moment $\langle x^{-1}\rangle_{\pi}$, an important integral characteristic of $\Phi_\pi$, is
\begin{equation}\label{eq:x^{-1}}
\langle x^{-1}\rangle_{\pi} \equiv \int_0^1\Phi_\pi(x;M) \frac{dx}{x} \approx 3.4 \text{ for }  M^2 \in [M^2_{-}, M^2_{+} ].
\end{equation}
This estimate of $\langle x^{-1}\rangle_{\pi}$ seems reasonable because the inverse moment is mostly formed by the behavior of DA near the left endpoint.
The estimate in (\ref{eq:x^{-1}}) is only a bit higher than the previous ones obtained in NLC SR \cite{Bakulev:2001pa} and lies within the acceptable region of the phenomenological analysis of the pion transition form factor (TFF) \cite{Bakulev:2003cs}.


\subsection{\hspace*{-4mm} Effects of renormalon-chain corrections to DA $\varphi^{\|}_{\rho}$}

For the case of $\rho_{\|}$ DA that is determined from the NLC SR in the vector channel, the 4-quark NLC contribution $\Phi_S$ comes with the opposite sign relative to the pion case, which leads to the relation \cite{Mikhailov:2021kty}
\begin{align}\label{eq:rho-pi-DAs}
\varphi_{\rho}^{\|}(x)
\approx
  \left[
        \varphi_\pi(x)
        - \frac{2}{f_{\pi}^2}\Phi_S(x,M)\!+\! \Delta_{A_{1}\rho'}(x, M) \right]
        e^{C(M)} ,
\end{align}
where
\begin{align}\label{eq:C}
C(M) = \frac{ m_{\rho}^2}{M^2} + \ln \left(f_{\pi}^2/f_{\rho}^2\right),
\end{align}
\begin{align}\label{eq:Delta-A1}
\Delta_{A_{1}\rho'}(x, M)
= {}&\left(\frac{f_{A_1}}{f_\pi} \right)^2 e^{-m_{A_1}^2/M^2} \varphi_{A_1}(x)
\notag\\
&{}- \left(\frac{f_{\rho'}}{f_\pi} \right)^2 e^{-m_{\rho'}^2/M^2} \varphi_{\rho'}(x).
\end{align}
The symbol ``approximately equal'' in Eq.~(\ref{eq:rho-pi-DAs}) means that we suppose $\Phi_\text{PT}(x;M;s_0^A) \approx \Phi_\text{PT}(x;M;s_0^V)$ for the purely perturbative parts in both channels. The term $\Delta_{A_{1}\rho'}$ is determined by the difference of the contributions of higher resonances in the phenomenological parts of QCD SR for the axial and vector channels.

\begin{figure}[h]
\begin{center}
\includegraphics[width=.47\textwidth]{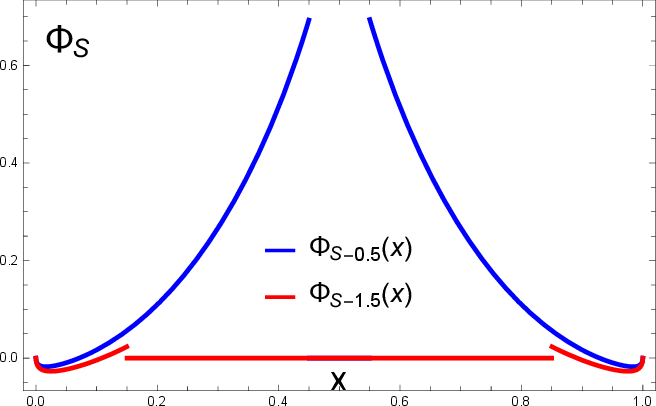}
\includegraphics[width=.47\textwidth]{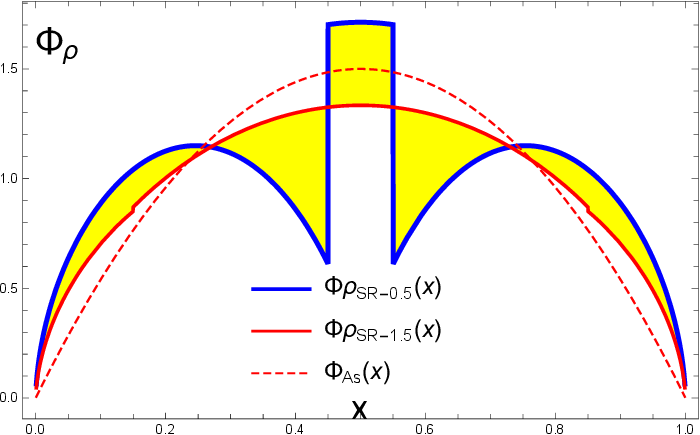}
\caption{\label{fig:qcdsr-rho}Top: NLC scalar condensate $\Phi_\text{S}$ up to $a_s(a_s\beta_0)^3$  for the parameter $M^2$ in the interval $[M^2_- = 0.5$ (blue line), $M^2_+ = 1.5$ (red line)] GeV$^2$.
Bottom: the rhs\ of NLC SR for $\varphi_{\rho}^{\|}$, Eq.~(\ref{eq:rho-pi-DAs}) up to $a_s(a_s\beta_0)^3$
in the interval $[M^2_-$,  $M^2_+]$ GeV$^2$.}
\end{center}
\end{figure}

Keeping only the contribution $\Phi_\text{PT}(x;M)-\Phi_S(x;M)$ (reduced NLC SR) in the rhs\ of (\ref{eq:sr-rho}), one gets the profile of $\varphi_{\rho}^{\|}(x)$ that becomes wider near the endpoints. This endpoint ``swelling'' of the $\varphi_{\rho}^{\|}(x)$ profile is seen clearly in Fig.~\ref{fig:qcdsr-rho} (bottom panel). This effect can be traced back also in the representation (\ref{eq:rho-pi-DAs}) for $\varphi_{\rho}^{\|}(x)$ through $\varphi_\pi(x)$.
The incline of the $\rho^{\|}$ meson DA near the endpoints is certainly larger than for the pion DA
and averages between 9 and 12 with the value of the inverse moment being $\langle x^{-1}\rangle_\rho \approx 3.8$.

Here it is impossible to reliably estimate the moments $\langle \xi^n\rangle$, $\xi = x -\bar{x}$ of $\pi$ and $\rho^{\|}$ DAs due to the fact that
the reduced NLC SR neglects some of the condensate contributions,
but we can still suggest inequalities for the moments.
The significant swelling effect near the endpoints should lead to the obvious inequalities
\begin{equation} \label{eq:ineq-xi2}
\langle \xi^2\rangle_{\rho^{\|}} > \langle \xi^2\rangle_{\pi}> \langle \xi^2\rangle_\text{As}=\frac{1}{5},
\end{equation}
and, therefore,
\begin{equation}\label{eq:ineq-a2}
a_2^{\rho^{\|}} > a_2^{\pi}>0,
\end{equation}
where $a_2^\text{M}$ is the 2nd Gegenbauer moment of DA of a meson M.
Since the omitted contribution $\Delta_C$ in the rhs\ of the NLC SR (\ref{eq:sr}) is the same for both channels, it does not violate the inequalities.
The results of lattice calculations \cite{Braun:2016wnx,Bali:2019dqc} support the conclusion \eqref{eq:ineq-a2}, 
\begin{gather}\label{eq:LQCD-a2}
a^{\rho^{\|}}_2=0.132(27) \text{ \cite{Braun:2016wnx}} > a^{\pi}_2=0.116(20) \text{ \cite{Bali:2019dqc}},
\\
\text{while}\,a^{\pi}_2 \text{ (LattQCD \cite{Bali:2019dqc}}) \approx a^{\pi}_2 \text{ (SR \cite{Mikhailov:2021znq}}) \notag
\end{gather}
at $\mu_\text{Latt}^2=4$ GeV$^2$, $\lambda^2_q \approx 0.45$ GeV$^2$.
Note that the previous versions of NLC SR for DAs of $\pi$ and $\rho^{\|}$ \cite{Mikhailov:2021kty,Pimikov:2013usa} yielded an opposite hierarchy of the moments, $a^{\rho^{\|}}_2 = 0.032(46) < a^{\pi}_2 = 0.149^{+0.052}_{-0.043}\,(\text{at}\, \mu^2\!\!=\!1$GeV$^2$).
The contributions of the orders $a_s(a_s\beta_0)^n$, $n=2$, 3  to the dominant components of SRs reverse this situation -- allowance for the renormalon-chain effects in the reduced NLC SRs of Eq.~(\ref{eq:sr}) provides a new estimate for $a^{\rho^{\|}}_2$ (in the same frame as for $a^\pi_2$) that complies with the hierarchy \eqref{eq:ineq-a2} suggested by lattice QCD:
$$
       a^{\rho^{\|}}_2 \approx 0.15 > a^{\pi}_2 \approx 0.07   \text{ at } \mu^2= 1\text{ GeV}^2.
$$
Let us emphasize that we insist on the validity of inequality \eqref{eq:ineq-a2} for $a_2^\text{M}$, $\text{M}= \pi$ and $\rho_{\|}$, per se, rather than on the precise values of the moments which serve \textit{only as an illustration here}.
To obtain well-grounded estimates of the moments $a_2^\text{M}$ one needs the standard treatment of the complete NLC SRs (\ref{eq:sr}).

\section{Conclusion}

(i) Taking into account renormalon-chain corrections of any order $a_s\left(a_s\beta_0\right)^n$ in pQCD, we have calculated the correlator $\Pi(x,\underline{0};L)$ of two vector quark currents, with one of the currents being nonlocal, which makes the correlator a function of the Bjorken fraction $x$.
The generating functions for this correlator and some of its moments have been constructed.
The zeroth and inverse Mellin moments of the correlator have been obtained for any order $n$.
The zeroth moment as well as some other fixed-order special cases agree with all previous calculations in the literature.

The correlator $\Pi(x,\underline{0};L)$ at any fixed order $n$ can be expressed in terms of harmonic polylogarithms of weight not higher than $n+3$.
Investigating the asymptotic series in $a_s\left(a_s\beta_0\right)^n$ for the moments of the correlator, we found
that these series should be truncated at $n=3$ or 4.

(ii)These radiative corrections to perturbative and condensate parts of QCD SR for pion distribution amplitude,
$\varphi_\pi(x)$, do not change the behavior of $\varphi_\pi(x)$ at the endpoints $x=0$ and 1 significantly.
Although these changes looks visible and corresponding incline is a bit higher now -- up to $7$. But this effect cannot  disturb the agreement of previous calculations of transition form factor and the phenomenological processing of the data  process $\gamma+\gamma^* \to \pi^0$ \cite{Bakulev:2003cs, Mikhailov:2021znq}.

(iii) The same class of radiative corrections to the distribution amplitude of longitudinally polarized $\rho$-meson, $\varphi_{\rho}^{\|}(x)$,
drastically changes the behavior of the DA near the endpoints in such a way that leads to the inequality $\langle \xi^2\rangle_{\rho^{\|}} > \langle\xi^2\rangle_{\pi}$ ($a_2^{\rho^{\|}} > a_2^{\pi}>0$). This inequality agrees with the results of lattice calculations in \cite{Braun:2016wnx,Bali:2019dqc}.


\appendix
\section{\label{sec:Borel}BOREL TRANSFORM $\mathbf{\hat{B}}$}
We used the standard form of the Borel transform \textit{for QCD SR}, see \eg in \cite{Shifman:1978bx}, it reads $\hat{\mathbf{B}}_{\scriptstyle{(M^2 \to P^2)}}[f(t)]$
and manifests itself as the limit of a series of derivatives of the function $f(t)$ for $t=P^2/\mu^2$
($\mu^2$ -- normalization scale)
\begin{gather}
\label{eq:borel}
\!\!\!\!\!\hat{\mathbf{B}}_{\scriptscriptstyle{(M^2 \to P^2)}}[f(t)]\!\equiv\!\mathbf{\hat{B}} \left[ f(t) \right]\!\!(M^2)\!\stackrel{def}{=}\!\!\lim_{\begin{subarray}{c} \scriptscriptstyle{P^2=n M^2} \\ \scriptscriptstyle{n\to \infty}\end{subarray}} \frac{(-t)^n}{\Gamma(n)} \frac{\mathrm d^n }{\mathrm d t^n} \left[f(t)\right].
\end{gather}
We emphasize that the Borel transformation $\mathbf{\hat{B}}_{\scriptscriptstyle{(M^2 \to P^2)}}$,
acts on the argument $P^2$, this
differs from the images of $\mathbf{\tilde B}$ (the inverse Laplace transform) acting on the powers
of $a_s$ (or the constant $A \sim a_s$) of the perturbation theory series,
the latter have been summed and discussed  here in Sec.\,\ref{sec:2.2},\,\ref{sec:2.3}.
A number of useful formulae for the $\hat{\mathbf{B}}$ are presented below that are based on the definition (\ref{eq:borel})
\begin{eqnarray} \label{eq:A2}
\!\!\!\!\mathbf{\hat{B}} \left[ \exp\left(- \frac{P^2}{\mu^2} a\right)\right]\!&=&\! \delta\left(1-\frac{M^2}{\mu^2} a\right).
\end{eqnarray}
 Based on (\ref{eq:A2}) one can derive 
\begin{subequations}
\begin{eqnarray}
\Rightarrow \mathbf{\hat{B}}\left[\left(\frac{\mu^2}{P^2}\right)^n\right]&=&\frac{1}{\Gamma(n)}\left(\frac{\mu^2}{M^2}\right)^n, \\
\mathbf{\hat{B}} \left[e^{a L}\right] &=& - \frac{a e^{a l}}{\Gamma(1-a)}.
\end{eqnarray}
\end{subequations}
Here $a$ is a constant, \eg, $a=A= -a_s\beta_0$ as in Sec.\ref{sec:2}, $L \!=\! \ln t$, $l\!=\!\ln\left(\frac{M^2}{\mu^2}\right)$.
The $\mathbf{\hat{B}}$--images of radiation logs are:
\begin{equation}
\mathbf{\hat{B}}~\frac{d}{d L} = \frac{d}{d l}~\mathbf{\hat{B}}~~~ \Rightarrow ~~~\mathbf{\hat{B}}\left(\frac{d}{d L}\Pi=\dot{\Pi}\right)=\frac{d}{d l}~\mathbf{\hat{B}}\Pi;
\end{equation}
\begin{subequations}
\begin{align}
\!\!\!\!\mathbf{\hat{B}} \left[ \ln^{m}(t) \right](M^2)\! =& m(-1)^m
\left[  \left( \frac{d}{da} \right)^{m-1}\!\!\!\! \frac{e^{-a l}}{\Gamma(1+a)}\right]_{a=0}  \\
\!\!\!\!=&\!\! - m \!  \left(l_B - \frac{d}{da} \right)^{m-1}\!\!\!\! \frac{e^{-\gamma_\text{E} a}}{\Gamma(1+a)}\bigg|_{a=0}\!\!;
\end{align}
\end{subequations}
here $l_B\!=\!\ln\left(\frac{M^2}{\mu^2}\right)-\gamma_\text{E}$.

\section{\label{sec:rho} EXTRACTION OF SPECTRAL DENSITY $\rho(s)$}
Let us define a ``double'' Borel transform $\mathbf{\hat{B}^2}_{(s)}\!\! \equiv \!\mathbf{\hat{B}^2}_{\scriptscriptstyle{(\!s \to P^2\!)}}$ to obtain the spectral density $\rho_\text{pt}(s)$ of $\Pi(L)$ that is used in Sec.\ref{sec:SRs} for QCD SR,
\begin{subequations}
\begin{gather}
\Pi(L) = \int^\infty_0 \frac{\rho_\text{pt}(s) ds}{s+P^2}-\text{subtracted terms},
\\
M^2\mathbf{\hat{B}}_{\scriptscriptstyle{(M^2\to P^2)}}\Pi(L) = \int^\infty_0 \rho_\text{pt}(s) e^{-s/M^2} ds,
\\
\mathbf{\hat{B}^2}_{(s)} \Pi(L)\!\equiv \!\frac{1}{s}\mathbf{\hat{B}_{(\frac{1}{s}\to \sigma)}}\!\left[\frac{1}{\sigma}\mathbf{\hat{B}_{(\frac{1}{\sigma}\to P^2)}}\Pi(L)\right]\!=\!\rho_\text{pt}(s),
\end{gather}
\end{subequations}
where $\sigma$ is an intermediate variable.
One obtains for every power $L^n$ in $\Pi(L)$ the contribution to $\rho_\text{pt}(s)$ as a polynomial in $l_s=\ln(s)$:
\begin{eqnarray}
\mathbf{\hat{B}^2}_{(s)} L^n&=& \left(l_s - \frac{d}{d\nu} \right)^n \left[ \frac{\sin(\pi \nu)}{\pi}\right]\bigg|_{\nu=0},\label{eq:B2} \\
\mathbf{\hat{B}^2}_{(s)}\left[ e^{a L} \right]&=& -\displaystyle e^{a l_s}\, \frac{\sin(\pi a)}{\pi}.
\end{eqnarray}
\begin{widetext}
\begin{ruledtabular}
\begin{tabular}{ccccccc}
	                      $n$& 0 &	1&        2& 3    & 4 & 5 \\ \hline
$\mathbf{\hat{B}^2}_{(s)}L^n$& 0 & $-1$& $-2 l_s$  & $\pi^2-3l_s^2$     &$4l_s\pi^2 -4l_s^3$ & $-\pi^4+10\pi^2 l_s^2 -5 l_s^4$
\end{tabular}
\end{ruledtabular}
\end{widetext}
The key element of the perturbative contribution in the ``theoretical part" (rhs) of the SR is the integration of $\ln^j(s)$ from (\ref{eq:B2}),
\begin{align}
\int^{s_{0}}_0 \rho_\text{pt}(s) e^{-s/M^2}ds \Rightarrow M^2 \int_0^{s_{0}/M^2}\ln^j(M^2t) e^{-t}dt\,.
\end{align}
Taking into account the main terms of the structure of the correlator after summation,
\ie, the terms of the generating functions for $\Pi'$ and $\Pi''$ in Secs.~\ref{sec:2}
(see the terms in  braces below),
we present the results for these functions and their different derivatives where $a$ is a constant,
\begin{widetext}
\begin{align}
\rho_\text{pt}(s)
&\propto \mathbf{\hat{B}^2}_{(s)} \Big\{ \frac{\exp{(a L)}}{a}, L  \Big \}
= \Big\{ - \exp{(a l_s)} \frac{\sin(\pi a)}{\pi a},\,-1\Big\}\,,
\\
\mathbf{\hat{B}}\Pi(L)
&\propto \mathbf{\hat{B}} \Big\{ \frac{\exp{(a L)}}{a},\, L  \Big \}
= \Big\{ - \frac{\exp{(a l)}}{\Gamma(1-a)},\,-1 \Big\}\,,
 \\
\dot{\Pi}(L)
&\propto \frac{d}{dL}\Big\{ \frac{\exp{(a L)}}{a}, L \Big\}
= \Big\{  \exp{(a L)},1 \Big\}.
\end{align}
\end{widetext}

\section{\label{sec:app.C}$\Phi_S$ INTEGRALS}

The zeroth-order calculation $a_s (a_s \beta_0)^0$ of the coefficient function for $\Phi_S $,discussed in detail
 in \cite{Mikhailov:1988nz} (see also Appendix~A in \cite{Mikhailov:2010ud}), was performed for the correlator of the initial two-fold form $\Pi_S(x,y)$.
For this two-fold form, the contribution $\Pi^{(n)}_S(x,y)$ with a $n$-bubble chain looks most evident as a term of geometric progression
\begin{widetext}
\begin{eqnarray}
\mathbf{\hat{B}}\Pi^{(n)}_S(x,y) &\sim & (a_s \beta_0)^n \mathop{\hat{\mathbf{S}}} f(x,y)
\left[ \ln\left(\frac{\bar{\Delta}}{y-x}-1\right)-\ln\left(\frac{\delta^2}{\mu_0^2} \right) -c \right]^n ; \\
         &&\Delta=\frac{\delta^2}{M^2};~ \bar{\Delta}=1-\Delta;~ \delta^2= \frac{\lambda_q^2}2;~\mu_0^2=\mu^2e^{\gamma_E}; ~c=-\frac{5}{3}; \\
  f(x,y) &=&\frac{16}{9}\pi\frac{\langle \sqrt{\alpha_s}\bar{q}q\rangle^2}{ \delta^4 \bar{\Delta}} \frac{\bar{x}y}{y-x+\Delta}\theta(\bar{\Delta}>y-x)\theta(y>\bar{\Delta})\theta(y>x)\theta(\Delta > x).
\end{eqnarray}
\end{widetext}
 Finally, we integrate over $y$ to obtain the contribution to $\Pi_S(x,\underline{0}) \sim \Phi_S(x)$.
 The partial contribution $\Phi^{(n)}_S$ of the order $(a_s \beta_0)^n$ to $\Phi_S$ reads
\begin{widetext}
\begin{eqnarray}
 \label{eq:Cfunc}
\Phi^{(n)}_S(x)&=&(a_s \beta_0)^n \frac{16}{9}\pi \frac{\langle \sqrt{\alpha_s}\bar{q}q\rangle^2}{\delta^4\bar{\Delta}}
\theta(1> 2\Delta)
\theta(\Delta >x) \bar{x}
\int^{\bar{\Delta}+x}_{\bar{\Delta}} \frac{y}{y-x+\Delta}
\left[ \ln\left(\frac{\bar{\Delta}}{y-x}-1\right)-\ln\left(\frac{\delta^2}{\mu_0^2} \right) -c \right]^n\, dy \nonumber\\
&&{}+(x\to \bar{x})
\end{eqnarray}
 \end{widetext}
The  functions of the highest transcendence that appear in  $\Phi^{(n)}_S(x)$ from Eq.~(\ref{eq:Cfunc}) are the polylogarithms of weight $n+1$,
\begin{equation}
\mathop{\mathrm{Li}_{n+1}}\left(-\frac{x}{\bar{\Delta}-x}\right),\mathop{\mathrm{Li}_{n+1}}
\left(-\frac{x \Delta}{\bar{\Delta}-x}\right),
(x\leftrightarrow \bar{x})
\end{equation}



\begin{thebibliography}{30}
\newcommand{\noopsort}[1]{} \newcommand{\printfirst}[2]{#1}
  \newcommand{\singleletter}[1]{#1} \newcommand{\switchargs}[2]{#2#1}
\expandafter\ifx\csname natexlab\endcsname\relax\def\natexlab#1{#1}\fi
\expandafter\ifx\csname bibnamefont\endcsname\relax
  \def\bibnamefont#1{#1}\fi
\expandafter\ifx\csname bibfnamefont\endcsname\relax
  \def\bibfnamefont#1{#1}\fi
\expandafter\ifx\csname citenamefont\endcsname\relax
  \def\citenamefont#1{#1}\fi
\expandafter\ifx\csname url\endcsname\relax
  \def\url#1{\texttt{#1}}\fi
\expandafter\ifx\csname urlprefix\endcsname\relax\def\urlprefix{URL }\fi
\providecommand{\bibinfo}[2]{#2}
\providecommand{\eprint}[2][]{\url{#2}}

\bibitem[{\citenamefont{Bakulev and Mikhailov}(1998)}]{Bakulev:1998pf}
\bibinfo{author}{\bibfnamefont{A.~P.} \bibnamefont{Bakulev}} \bibnamefont{and}
  \bibinfo{author}{\bibfnamefont{S.~V.} \bibnamefont{Mikhailov}},
  \bibinfo{journal}{Phys. Lett.} \textbf{\bibinfo{volume}{B436}},
  \bibinfo{pages}{351} (\bibinfo{year}{1998}), \eprint{hep-ph/9803298}.

\bibitem[{\citenamefont{Mikhailov and Stefanis}(2021)}]{Mikhailov:2021kty}
\bibinfo{author}{\bibfnamefont{S.~V.} \bibnamefont{Mikhailov}}
  \bibnamefont{and} \bibinfo{author}{\bibfnamefont{N.~G.}
  \bibnamefont{Stefanis}}, \bibinfo{journal}{Phys. Rev. D}
  \textbf{\bibinfo{volume}{104}}, \bibinfo{pages}{096013}
  (\bibinfo{year}{2021}), \eprint{2106.15522}.

\bibitem[{\citenamefont{Mikhailov and
  Volchanskiy}(2020{\natexlab{a}})}]{Mikhailov:2020izb}
\bibinfo{author}{\bibfnamefont{S.~V.} \bibnamefont{Mikhailov}}
  \bibnamefont{and}
  \bibinfo{author}{\bibfnamefont{N.}~\bibnamefont{Volchanskiy}},
  \bibinfo{journal}{J. Phys. Conf. Ser.} \textbf{\bibinfo{volume}{1435}},
  \bibinfo{pages}{012059} (\bibinfo{year}{2020}{\natexlab{a}}).

\bibitem[{\citenamefont{Mikhailov and Volchanskiy}(2023)}]{Mikhailov:2023lqe}
\bibinfo{author}{\bibfnamefont{S.~V.} \bibnamefont{Mikhailov}}
  \bibnamefont{and}
  \bibinfo{author}{\bibfnamefont{N.}~\bibnamefont{Volchanskiy}},
  \bibinfo{journal}{Phys. Part. Nuclei Lett.} \textbf{\bibinfo{volume}{20}},
  \bibinfo{pages}{296–299} (\bibinfo{year}{2023}), \eprint{2301.01806}.

\bibitem[{\citenamefont{Broadhurst and Grozin}(1995)}]{Broadhurst:1994se}
\bibinfo{author}{\bibfnamefont{D.~J.} \bibnamefont{Broadhurst}}
  \bibnamefont{and} \bibinfo{author}{\bibfnamefont{A.~G.}
  \bibnamefont{Grozin}}, \bibinfo{journal}{Phys. Rev. D}
  \textbf{\bibinfo{volume}{52}}, \bibinfo{pages}{4082} (\bibinfo{year}{1995}),
  \eprint{hep-ph/9410240}.

\bibitem[{\citenamefont{Mikhailov and Volchanskiy}(2019)}]{Mikhailov:2018udp}
\bibinfo{author}{\bibfnamefont{S.~V.} \bibnamefont{Mikhailov}}
  \bibnamefont{and}
  \bibinfo{author}{\bibfnamefont{N.}~\bibnamefont{Volchanskiy}},
  \bibinfo{journal}{JHEP} \textbf{\bibinfo{volume}{01}}, \bibinfo{eid}{202}
  (\bibinfo{year}{2019}), \eprint{1812.02164}.

\bibitem[{\citenamefont{Mikhailov}(1998)}]{Mikhailov:1998xi}
\bibinfo{author}{\bibfnamefont{S.~V.} \bibnamefont{Mikhailov}},
  \bibinfo{journal}{Phys. Lett. B} \textbf{\bibinfo{volume}{431}},
  \bibinfo{pages}{387} (\bibinfo{year}{1998}), \eprint{hep-ph/9804263}.

\bibitem[{\citenamefont{Mikhailov}(2000)}]{Mikhailov:1999ht}
\bibinfo{author}{\bibfnamefont{S.~V.} \bibnamefont{Mikhailov}},
  \bibinfo{journal}{Phys. Rev.} \textbf{\bibinfo{volume}{D62}},
  \bibinfo{pages}{034002} (\bibinfo{year}{2000}), \eprint{hep-ph/9910389}.

\bibitem[{\citenamefont{Mikhailov and
  Volchanskiy}(2020{\natexlab{b}})}]{Mikhailov:2020tta}
\bibinfo{author}{\bibfnamefont{S.~V.} \bibnamefont{Mikhailov}}
  \bibnamefont{and}
  \bibinfo{author}{\bibfnamefont{N.}~\bibnamefont{Volchanskiy}},
  \bibinfo{journal}{JHEP} \textbf{\bibinfo{volume}{21}}, \bibinfo{pages}{197}
  (\bibinfo{year}{2020}{\natexlab{b}}), \eprint{2010.03557}.

\bibitem[{\citenamefont{Remiddi and Vermaseren}(2000)}]{Remiddi:1999ew}
\bibinfo{author}{\bibfnamefont{E.}~\bibnamefont{Remiddi}} \bibnamefont{and}
  \bibinfo{author}{\bibfnamefont{J.~A.~M.} \bibnamefont{Vermaseren}},
  \bibinfo{journal}{Int. J. Mod. Phys. A} \textbf{\bibinfo{volume}{15}},
  \bibinfo{pages}{725} (\bibinfo{year}{2000}), \eprint{hep-ph/9905237}.

\bibitem[{\citenamefont{Laenen et~al.}(2023)\citenamefont{Laenen, Marinissen,
  and Vonk}}]{Laenen:2023hzu}
\bibinfo{author}{\bibfnamefont{E.}~\bibnamefont{Laenen}},
  \bibinfo{author}{\bibfnamefont{C.}~\bibnamefont{Marinissen}},
  \bibnamefont{and} \bibinfo{author}{\bibfnamefont{M.}~\bibnamefont{Vonk}}
  (\bibinfo{year}{2023}), \eprint{2302.13715}.

\bibitem[{\citenamefont{Ball et~al.}(1995)\citenamefont{Ball, Beneke, and
  Braun}}]{Ball:1995ni}
\bibinfo{author}{\bibfnamefont{P.}~\bibnamefont{Ball}},
  \bibinfo{author}{\bibfnamefont{M.}~\bibnamefont{Beneke}}, \bibnamefont{and}
  \bibinfo{author}{\bibfnamefont{V.~M.} \bibnamefont{Braun}},
  \bibinfo{journal}{Nucl. Phys. B} \textbf{\bibinfo{volume}{452}},
  \bibinfo{pages}{563} (\bibinfo{year}{1995}), \eprint{hep-ph/9502300}.

\bibitem[{\citenamefont{{Broadhurst}}(1993)}]{1993ZPhyC..58..339B}
\bibinfo{author}{\bibfnamefont{D.~J.} \bibnamefont{{Broadhurst}}},
  \bibinfo{journal}{Zeitschrift f\"ur Physik C} \textbf{\bibinfo{volume}{58}},
  \bibinfo{pages}{339} (\bibinfo{year}{1993}).

\bibitem[{\citenamefont{Broadhurst and Kataev}(1993)}]{Broadhurst:1993ru}
\bibinfo{author}{\bibfnamefont{D.~J.} \bibnamefont{Broadhurst}}
  \bibnamefont{and} \bibinfo{author}{\bibfnamefont{A.~L.}
  \bibnamefont{Kataev}}, \bibinfo{journal}{Phys. Lett. B}
  \textbf{\bibinfo{volume}{315}}, \bibinfo{pages}{179} (\bibinfo{year}{1993}),
  \eprint{hep-ph/9308274}.

\bibitem[{\citenamefont{Mikhailov and Radyushkin}(1989)}]{Mikhailov:1988nz}
\bibinfo{author}{\bibfnamefont{S.~V.} \bibnamefont{Mikhailov}}
  \bibnamefont{and} \bibinfo{author}{\bibfnamefont{A.~V.}
  \bibnamefont{Radyushkin}}, \bibinfo{journal}{Sov. J. Nucl. Phys.}
  \textbf{\bibinfo{volume}{49}}, \bibinfo{pages}{494} (\bibinfo{year}{1989}),
  \bibinfo{note}{{Yad. Fiz. 49, 794 (1988), JINR-P2-88-103 (in Russian)}},
  \urlprefix\url{http://inspirehep.net/record/262441/files/JINR-P2-88-103.pdf}.

\bibitem[{\citenamefont{Shifman et~al.}(1979)\citenamefont{Shifman, Vainshtein,
  and Zakharov}}]{Shifman:1978bx}
\bibinfo{author}{\bibfnamefont{M.~A.} \bibnamefont{Shifman}},
  \bibinfo{author}{\bibfnamefont{A.~I.} \bibnamefont{Vainshtein}},
  \bibnamefont{and} \bibinfo{author}{\bibfnamefont{V.~I.}
  \bibnamefont{Zakharov}}, \bibinfo{journal}{Nucl. Phys.}
  \textbf{\bibinfo{volume}{B147}}, \bibinfo{pages}{385} (\bibinfo{year}{1979}).

\bibitem[{\citenamefont{Ball and Braun}(1996)}]{PhysRevD.54.2182}
\bibinfo{author}{\bibfnamefont{P.}~\bibnamefont{Ball}} \bibnamefont{and}
  \bibinfo{author}{\bibfnamefont{V.~M.} \bibnamefont{Braun}},
  \bibinfo{journal}{Phys. Rev. D} \textbf{\bibinfo{volume}{54}},
  \bibinfo{pages}{2182} (\bibinfo{year}{1996}),
  \urlprefix\url{https://link.aps.org/doi/10.1103/PhysRevD.54.2182}.

\bibitem[{\citenamefont{Bakulev et~al.}(2001)\citenamefont{Bakulev, Mikhailov,
  and Stefanis}}]{Bakulev:2001pa}
\bibinfo{author}{\bibfnamefont{A.~P.} \bibnamefont{Bakulev}},
  \bibinfo{author}{\bibfnamefont{S.~V.} \bibnamefont{Mikhailov}},
  \bibnamefont{and} \bibinfo{author}{\bibfnamefont{N.~G.}
  \bibnamefont{Stefanis}}, \bibinfo{journal}{Phys. Lett.}
  \textbf{\bibinfo{volume}{B508}}, \bibinfo{pages}{279} (\bibinfo{year}{2001}),
  \bibinfo{note}{[Erratum: Phys. Lett. B590, 309 (2004)]},
  \eprint{hep-ph/0103119}.

\bibitem[{\citenamefont{Bakulev
  et~al.}(2004{\natexlab{a}})\citenamefont{Bakulev, Mikhailov, and
  Stefanis}}]{Bakulev:2004mc}
\bibinfo{author}{\bibfnamefont{A.~P.} \bibnamefont{Bakulev}},
  \bibinfo{author}{\bibfnamefont{S.~V.} \bibnamefont{Mikhailov}},
  \bibnamefont{and} \bibinfo{author}{\bibfnamefont{N.~G.}
  \bibnamefont{Stefanis}}, \bibinfo{journal}{Annalen Phys.}
  \textbf{\bibinfo{volume}{13}}, \bibinfo{pages}{629}
  (\bibinfo{year}{2004}{\natexlab{a}}), \eprint{hep-ph/0410138}.

\bibitem[{\citenamefont{Mikhailov and Radyushkin}(1986)}]{Mikhailov:1986be}
\bibinfo{author}{\bibfnamefont{S.~V.} \bibnamefont{Mikhailov}}
  \bibnamefont{and} \bibinfo{author}{\bibfnamefont{A.~V.}
  \bibnamefont{Radyushkin}}, \bibinfo{journal}{JETP Lett.}
  \textbf{\bibinfo{volume}{43}}, \bibinfo{pages}{712} (\bibinfo{year}{1986}),
  \bibinfo{note}{[Pisma Zh. Eksp. Teor. Fiz. 43, 551 (1986)]}.

\bibitem[{\citenamefont{Mikhailov et~al.}(2021)\citenamefont{Mikhailov,
  Pimikov, and Stefanis}}]{Mikhailov:2021znq}
\bibinfo{author}{\bibfnamefont{S.~V.} \bibnamefont{Mikhailov}},
  \bibinfo{author}{\bibfnamefont{A.~V.} \bibnamefont{Pimikov}},
  \bibnamefont{and} \bibinfo{author}{\bibfnamefont{N.~G.}
  \bibnamefont{Stefanis}}, \bibinfo{journal}{Phys. Rev. D}
  \textbf{\bibinfo{volume}{103}}, \bibinfo{pages}{096003}
  (\bibinfo{year}{2021}), \eprint{2101.12661}.

\bibitem[{\citenamefont{Mikhailov et~al.}(2010)\citenamefont{Mikhailov,
  Pimikov, and Stefanis}}]{Mikhailov:2010ud}
\bibinfo{author}{\bibfnamefont{S.~V.} \bibnamefont{Mikhailov}},
  \bibinfo{author}{\bibfnamefont{A.~V.} \bibnamefont{Pimikov}},
  \bibnamefont{and} \bibinfo{author}{\bibfnamefont{N.~G.}
  \bibnamefont{Stefanis}}, \bibinfo{journal}{Phys. Rev.}
  \textbf{\bibinfo{volume}{D82}}, \bibinfo{pages}{054020}
  (\bibinfo{year}{2010}), \eprint{1006.2936}.

\bibitem[{\citenamefont{Bakulev
  et~al.}(2004{\natexlab{b}})\citenamefont{Bakulev, Mikhailov, and
  Stefanis}}]{Bakulev:2003cs}
\bibinfo{author}{\bibfnamefont{A.~P.} \bibnamefont{Bakulev}},
  \bibinfo{author}{\bibfnamefont{S.~V.} \bibnamefont{Mikhailov}},
  \bibnamefont{and} \bibinfo{author}{\bibfnamefont{N.~G.}
  \bibnamefont{Stefanis}}, \bibinfo{journal}{Phys. Lett.}
  \textbf{\bibinfo{volume}{B578}}, \bibinfo{pages}{91}
  (\bibinfo{year}{2004}{\natexlab{b}}), \eprint{hep-ph/0303039}.


\bibitem[{\citenamefont{Braun et~al.}(2017)}]{Braun:2016wnx}
\bibinfo{author}{\bibfnamefont{V.~M.} \bibnamefont{Braun}}
  \bibnamefont{et~al.}, \bibinfo{journal}{JHEP} \textbf{\bibinfo{volume}{04}},
  \bibinfo{pages}{082} (\bibinfo{year}{2017}), \eprint{1612.02955}.

\bibitem[{\citenamefont{Bali et~al.}(2019)\citenamefont{Bali, Braun,
  B{\"u}rger, G{\"o}ckeler, Gruber, Hutzler, Korcyl, Sch{\"a}fer, Sternbeck,
  and Wein}}]{Bali:2019dqc}
\bibinfo{author}{\bibfnamefont{G.~S.} \bibnamefont{Bali}},
  \bibinfo{author}{\bibfnamefont{V.~M.} \bibnamefont{Braun}},
  \bibinfo{author}{\bibfnamefont{S.}~\bibnamefont{B{\"u}rger}},
  \bibinfo{author}{\bibfnamefont{M.}~\bibnamefont{G{\"o}ckeler}},
  \bibinfo{author}{\bibfnamefont{M.}~\bibnamefont{Gruber}},
  \bibinfo{author}{\bibfnamefont{F.}~\bibnamefont{Hutzler}},
  \bibinfo{author}{\bibfnamefont{P.}~\bibnamefont{Korcyl}},
  \bibinfo{author}{\bibfnamefont{A.}~\bibnamefont{Sch{\"a}fer}},
  \bibinfo{author}{\bibfnamefont{A.}~\bibnamefont{Sternbeck}},
  \bibnamefont{and} \bibinfo{author}{\bibfnamefont{P.}~\bibnamefont{Wein}},
  \bibinfo{journal}{JHEP} \textbf{\bibinfo{volume}{08}}, \bibinfo{pages}{065}
  (\bibinfo{year}{2019}), \bibinfo{journal}{JHEP} \textbf{\bibinfo{volume}{11}}, \bibinfo{pages}{037}
  (\bibinfo{year}{2020}),
  \eprint{1903.08038}.

\bibitem[{\citenamefont{Pimikov et~al.}(2014)\citenamefont{Pimikov, Mikhailov,
  and Stefanis}}]{Pimikov:2013usa}
\bibinfo{author}{\bibfnamefont{A.~V.} \bibnamefont{Pimikov}},
  \bibinfo{author}{\bibfnamefont{S.~V.} \bibnamefont{Mikhailov}},
  \bibnamefont{and} \bibinfo{author}{\bibfnamefont{N.~G.}
  \bibnamefont{Stefanis}}, \bibinfo{journal}{Few Body Syst.}
  \textbf{\bibinfo{volume}{55}}, \bibinfo{pages}{401} (\bibinfo{year}{2014}),
  \eprint{1312.2776}.

\end{thebibliography}
\end{document}